\documentclass[aps,prl,twocolumn,preprintnumbers,groupedaddress,nofootinbib]{revtex4}

\usepackage{amsmath,color} 
\usepackage{graphicx} 
\usepackage{multirow}
\begin{document}

\preprint{FTUV-17-05-31}
\preprint{IFIC/17-31}

\title{Cosmology and CPT violating neutrinos}

\author{Gabriela Barenboim$^1$}
\email[]{Gabriela.Barenboim@uv.es}

\author{Jordi Salvado$^1$}
\email[]{jsalvado@ific.uv.es}

\affiliation{$^1$ Departament de F\'isica Te\`orica and Instituto de
  F\'isica Corpuscular,\\ Universitat de Val\`encia-CSIC, E-46100, Burjassot, Spain}

\date{\today}

\begin{abstract}
The combination Charge Conjugation-Parity-Time Reversal(CPT) is a fundamental symmetry in our current
understanding of nature. As such, testing CPT violation is a strongly
motivated path to explore new physics.  
In this paper we study CPT violation in the neutrino sector, giving
for the first time a bound, for a fundamental particle, in the CPT violating
particle-antiparticle gravitational mass difference. We argue that
cosmology is nowadays the only data sensitive to CPT violation for
the neutrino-antineutrino mass splitting and we use the latest
data release from Planck combined with the current
Baryonic-Acoustic-Oscillation measurement to perform a full
cosmological analysis. To show the potential of the future experiments 
we also show the results for Euclid, a next generation large scale structure
experiment.

\end{abstract}
\maketitle
\section{Introduction}

On general grounds, local, relativistic  quantum field theory makes
only a couple of predictions. CPT invariance \cite{CPT} is one of
them, and undoubtedly the cornerstone of our model building strategy.    
The CPT Theorem, in short, states that every  particle does have the
same mass as its anti-particle and, if unstable, also the same
lifetime.  Its position as one of the sacred cows of particle physics
is based on the fact that in order to prove it only three ingredients
are needed, all of which are "natural" and have other reasons to be in
our theory, way beyond the CPT theorem itself. They are 
\begin{itemize}
\item Lorentz invariance,
\item hermiticity of the Hamiltonian,
\item locality.
\end{itemize}
Precisely because of this, if CPT is found not to be conserved, the
impact of such an observation to fundamental physics would be
gigantic. It would necessarily mean that at least one of the three
assumptions above must be violated \cite{BL,Greenberg}. And therefore
it will automatically imply that our description of nature in terms of
local, Lorentz invariant field theory would be dramatically challenged
and our model building strategy would need to be seriously revisited. 

Largely because of its huge potential implications, the experimental
signature of CPT violation was searched in the past and according to
the PDG\cite{PDG}, the most  stringent limit on it comes from the neutral kaon
system\cite{Schwingenheuer:1995uf}.  Courtesy of the mixing between $K^0$ and $\overline{K}^0$,
the limit on the possible mass difference between them is beyond solid:
\begin{equation}
  \frac{|m(K^0) - m(\overline{K}^0)|}{m_K} < 0.6 \times 10^{-18} \label{eq:mK}
\end{equation}
However it is important to notice that the robustness of the CPT limit
from the neutral kaon system is somehow misleading. Although it is
nice to have a limit in a dimensionless way, we do not have a concrete
theory of CPT violation and therefore the scale with which we are
comparing the mass difference, the kaon mass in this case, is in every
way, arbitrary. A much stringent limit could have been obtained by
using the Planck mass instead, making exactly the same sense as the
one we currently use. \footnote{ Some  authors argue that the
  appropriate quantity to compare with $|m(K^0) - m(\overline{K}^0)|$
  in the analysis is $\Delta  m^2/E$ \cite{BBM}, although it is not
  evident why the merit of the bound should depend on the energy}

Until we have a full theory on CPT violation, the limit in
Eq.~(\ref{eq:mK}) should be looked as 
\begin{equation}
|m(K^0) - m(\overline{K}^0)| <
0.6 \times 10^{-18} m_K \simeq 10^{-9} \mbox{eV}.
\end{equation} 

Even more, as for bosons, the parameter entering the Lagrangian is the
mass squared, rather than the mass, the bound can alternative be written as
$|m^2(K^0) - m^2(\overline{K}^0)| < 0.25~\mbox{eV}^2 \label{eq:mK2}$,
which does not look nearly as strong as before. Besides, given that
the mass of the kaons is largely due to QCD, this test, cannot tell
directly whether elementary particles indeed respect the CPT
symmetry. For such a test, a search for CPT violation in the leptonic
sector is mandatory. Using charged leptons the most stringent bound comes from 
electron-positron $g-2$ experiments\cite{Dehmelt:1999jh, Bluhm:1997ci} and Hydrogen
Anti-Hydrogen spectroscopy\cite{Bluhm:1998rk}. These measurements
however, involve some combination between mass and charge as the testing parameter. On the other hand, in the neutral sector,
the discovery of neutrino oscillations established
that neutrinos are massive particles and in the so called See-Saw
models the light masses are naturally related with the grand 
unified scale making neutrinos distinctively sensitive to new physics/new scales. This exclusive mass generation mechanism along with the fact that there is no charge contamination comprised in the test makes
neutrinos specially appealing to study CPT violation. 

\begin{figure}[th!]
\begin{center}
\includegraphics[width=0.5\textwidth]{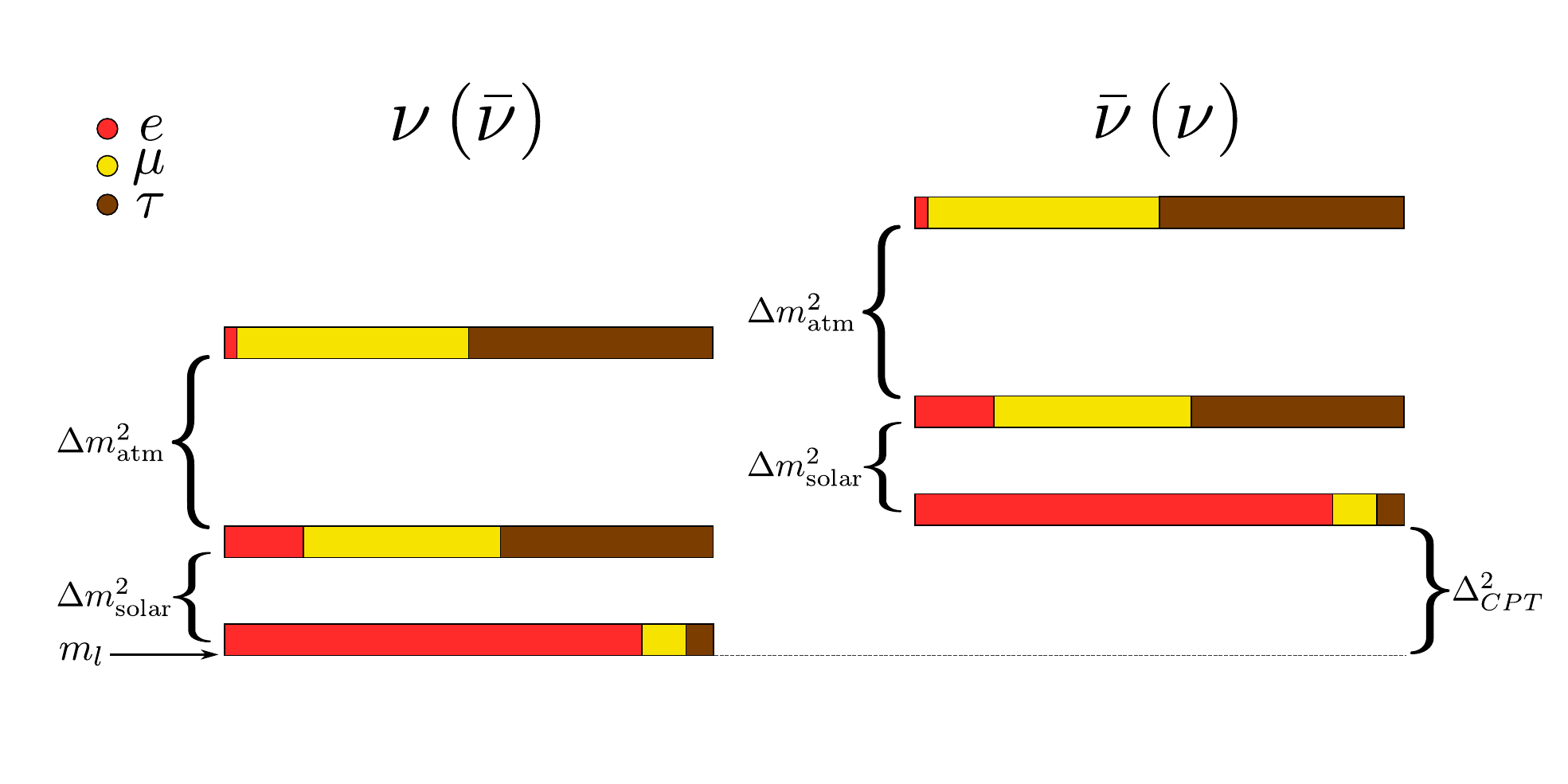}
\caption{Scheme for normal ordering mass spectrum with neutrinos and antineutrinos where we
  illustrate the extended parameters $\Delta_{CPT}$ and $m_l$.}
\label{spectrum}
\end{center}
\end{figure}

The quantum interference phenomena observed in neutrino oscillation is
very sensitive to new physics, and it has been proposed to
constrain CPT and Lorentz violation\cite{Barger:2000iv} in
solar\cite{BBM}, short and long base line\cite{Barenboim}, atmospheric
neutrino\cite{Abe:2014wla,Abbasi:2010kx} 
oscillations experiments. A constrain in full decoherent
oscillation regime using the recent discovered ultra high energy
neutrinos by IceCube\cite{Aartsen:2013jdh} has also been proposed
\cite{Arguelles:2015dca,Bustamante:2015waa}. 
In general neutrino oscillation physics has shown a strong potential
to constraint CPT being comparable or even stronger than that in the kaon
system\cite{BL2}.  

Unfortunately, all the experiments mentioned above always measure
$\Delta m^2$, and cannot measure the value of the masses themselves,
therefore, only CPT violation in the mass differences, {\it i.e.} $ \Delta
m_\nu^2 - \Delta m_{\bar{\nu}}^2$ can be tested.
Even more, if the possible violation of CPT has its origin in 
quantum gravity, we would naturally expect it to appear in the masses
themselves and not in the mass differences\cite{Fujikawa:2016her}. 

Here we focus on the study of the yet unconstrained CPT violating
mass difference between neutrinos and antineutrinos, $\Delta_{\rm
  CPT}=|m_l^\nu-m_l^{\bar \nu}|$.
It is worth noting, nevertheless, that the direct (kinematical)
searches for neutrino masses, carried out in tritium $\beta$-decay
experiments \cite{Dragoun:2015fqi} involve only anti(electron)
neutrinos and therefore strictly speaking only bound the masses in the
antineutrino sector, not probing anything about the neutrino one. An overall
shift on the spectrum, as the one shown in fig.\ref{spectrum}, which
can potentially be much larger than the mass differences themselves,
cannot be detected in neutrino oscillation experiments or bounded by future
direct kinematical searches. This leaves us the only option of using cosmological data for such a purpose.

In this article we give the first bound on CPT violation
for the neutrino-antineutrino absolute mass difference 
$\Delta_{\rm  CPT}$ using current cosmological
data. We also perform a forecast analysis for future next generation
European Space Agency Cosmic Vision mission, Euclid\cite{Laureijs:2011gra}\footnote{http://www.euclid-ec.org/}. 

\section{Cosmological Bounds}
Currently cosmology gives the strongest bound on the neutrino mass
scale. In the standard cosmological scenario 
neutrinos are produced thermally therefore, since neutrinos decouple
when they are relativistic, the number density for $\nu$ and
$\bar\nu$ in the cosmic neutrino background is the same. 
This implies that cosmology is giving a bound in neutrinos and
anti-neutrinos separately and therefore is currently the only physical
observable to both neutrino and anti-neutrino mass scales. 
Note that since gravitational interactions can not
distinguish particles from antiparticles, cosmology can only constrain
the absolute value of the mass difference and have no say on which spectrum is the heaviest/lightest. 

In this section we perform a Bayesian analysis for different sets
of cosmological observables. The cosmological model is given by
$\Lambda$CDM$+m_l+\Delta_{\rm cpt}$ where $\Lambda$CDM stands for the 6
standard cosmological parameters, $m_l$ for the value of the 
lightest neutrino mass and $\Delta_{\rm CPT}=|m_l^\nu-m_l^{\bar
 \nu}|$ is the absolute mass difference between neutrinos and
anti-neutrinos. The list of the cosmological parameters and the assumed
ranges in the analysis are given in table \ref{tab:parameters}. An
extra 94 fast sampling nuisance parameters are included to account for
systematic and calibration errors for Planck data\cite{Ade:2015xua},
in the case of Euclid forecast an extra nuisance parameter is
included\cite{Laureijs:2011gra}.

\begin{table}[h!]
\begin{tabular}{|c|c|}
\hline
Parameter & Prior\\
\hline
\hline
$\Omega_b h^2$ & $[0.001,0.1]$ \\
\hline
$\Omega _c h^2$ & $[0.01,0.99]$ \\
\hline
$100\Theta_s$ & $[0.01,10]$ \\
\hline
$n_s$ & $[0.5,1.5]$ \\
\hline
$\log(10^{10}A_s)$ & $[1,5]$ \\
\hline
$m_l$ (eV) & $[0,10]$\\
\hline
$\Delta_{CPT}$ (eV) & $[0,10]$\\
\hline
\end{tabular}
\caption{$\Lambda$CDM+$\nu$CPT parameters and the given ranges in
  where we take flat priors.}
\label{tab:parameters}
\end{table}

\begin{figure}
  \includegraphics[width=0.4\textwidth]{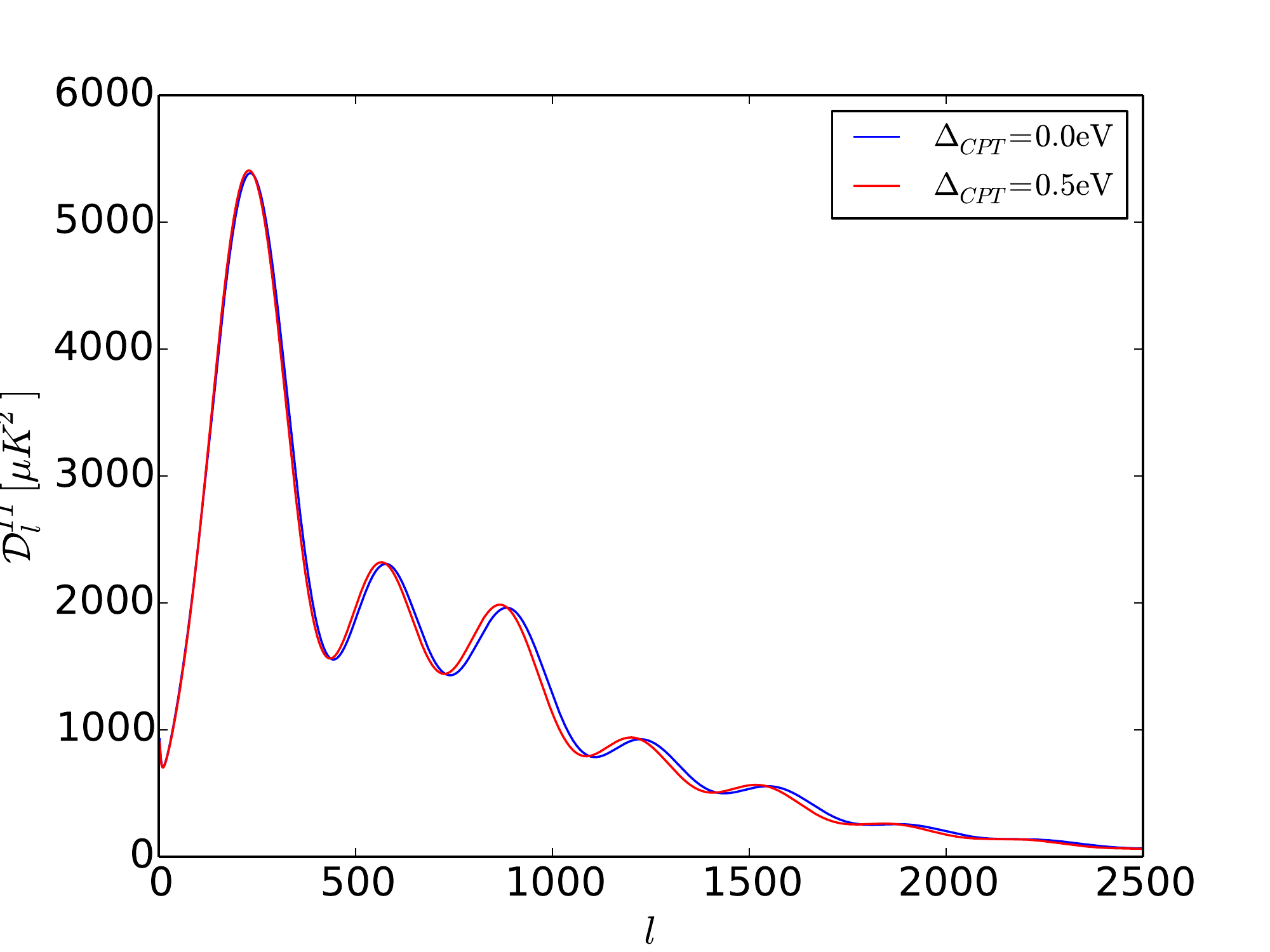}
  \includegraphics[width=0.4\textwidth]{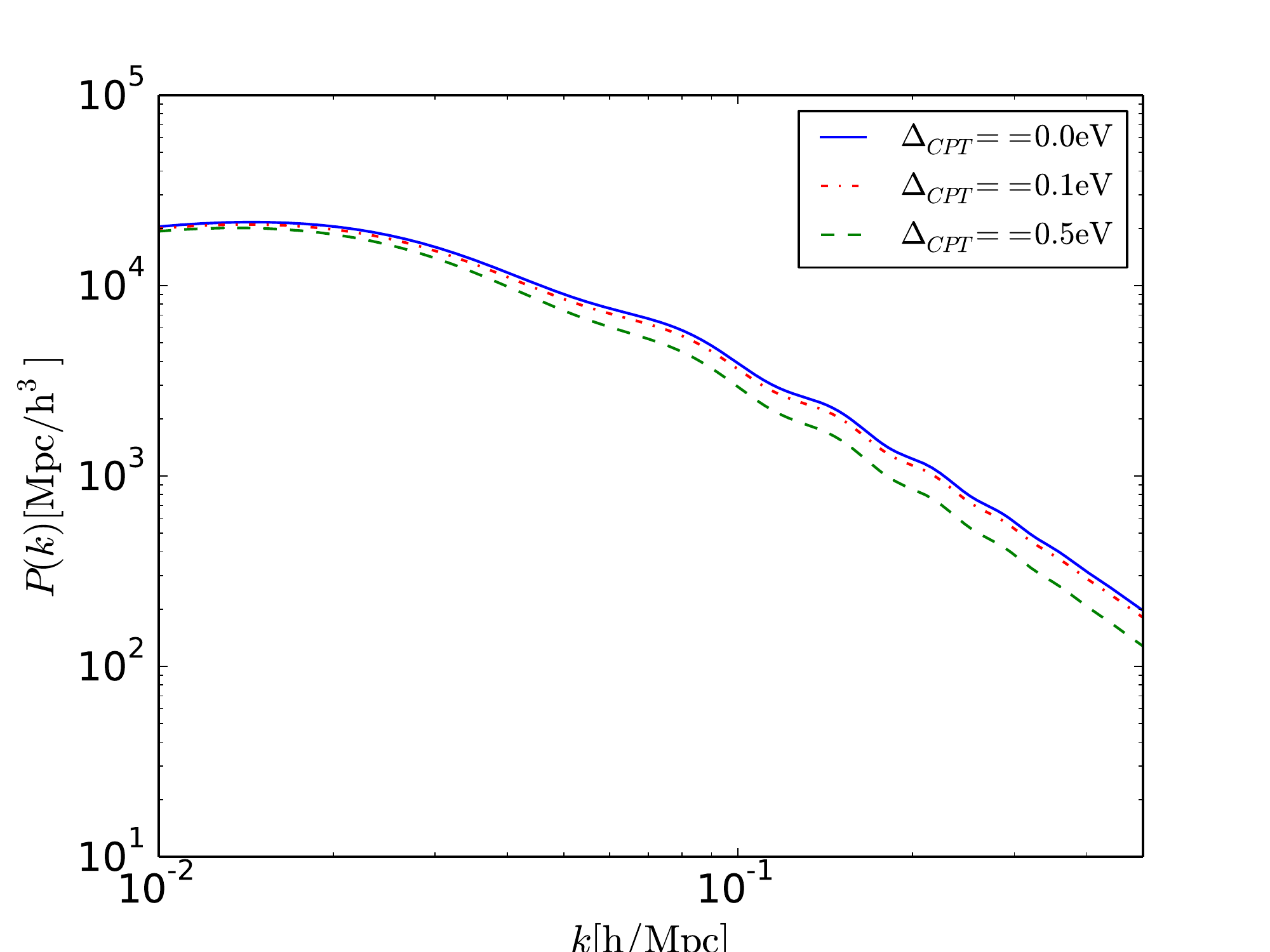}
  \caption{Effect of changing $\Delta_{CPT}$ in the TT CMB power 
      spectrum (top) and total matter power spectrum (bottom)}
\label{fig:effect}
\end{figure}

The effect of the neutrino masses in cosmology comes mainly via the free
streaming of the neutrinos in the cosmic neutrinos background during
the growth of the large scale structure.  
In fig.\ref{fig:effect} we show the effect in the
temperature-temperature (TT) CMB power spectrum and in the total matter power
spectrum for different values of the CPT violating mass splitting
$\Delta_{\rm CPT}$ and $m_l=0$., the rest of the cosmological
parameters are set to the Plank2015 $\Lambda CDM$ best fit\cite{Ade:2015xua}.

To perform the cosmological analysis we modify  by adding the new abovementioned parameters the publicly available
Boltzmann code CLASS\cite{Lesgourgues:2011re} to compute the linear
evolution of the cosmological perturbations and the MontePython
wrapper \cite{Audren:2012wb} to perform a Bayesian data analysis on
the full set of eight cosmological parameters. 

For both neutrinos and anti-neutrinos we fix the atmospheric and solar
mass splitting to the value of the global neutrino oscillation results
given by $\nu$-fit collaboration\footnote{http://www.nu-fit.org}\cite{Esteban:2016qun}
and we introduce the proper modifications to use $m_l$ and $\Delta_{\rm CPT}$ to parametrize 
the massive neutrinos.

\begin{figure}[ht!]
    \includegraphics[width=0.4\textwidth]{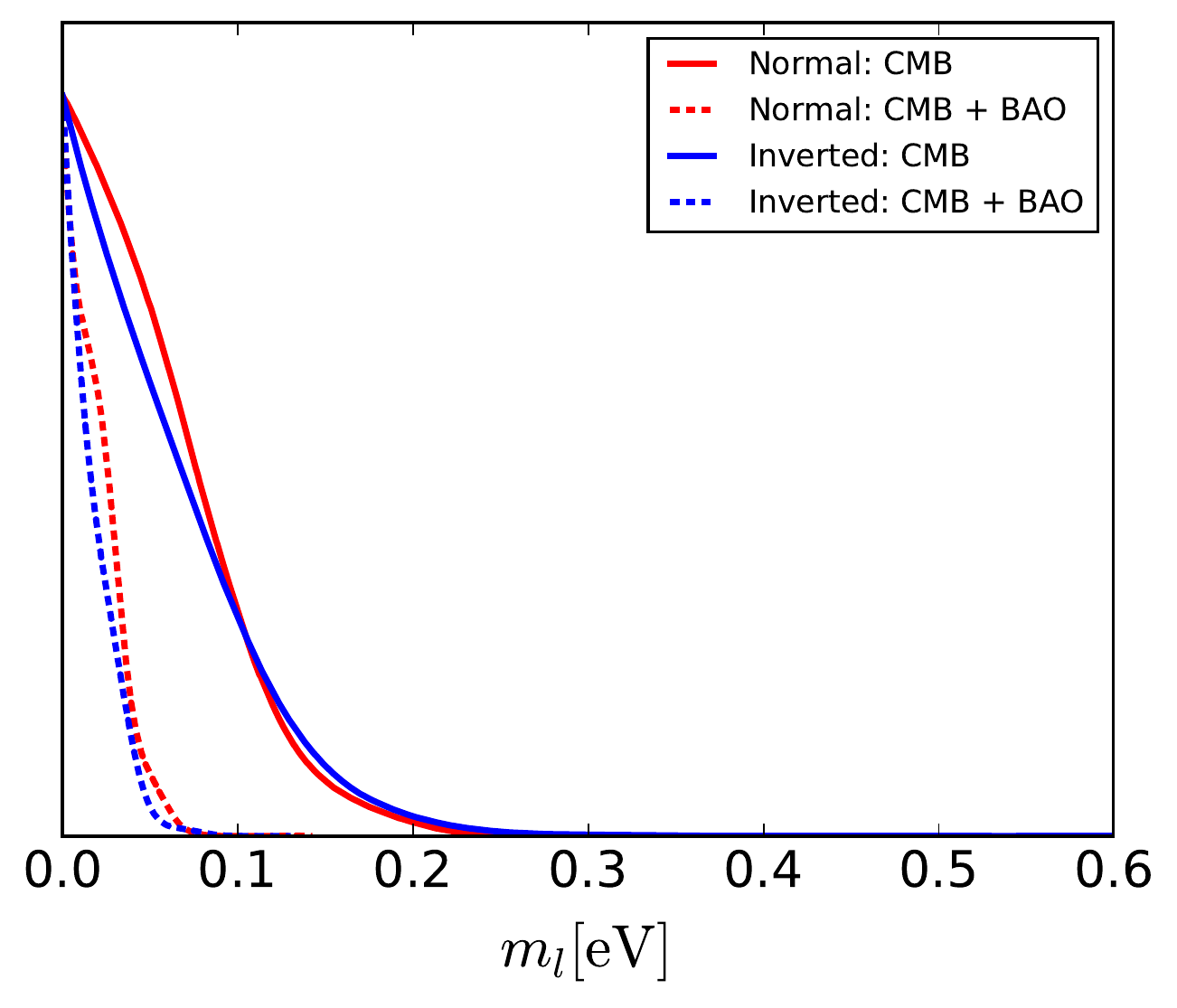}    
\caption{1D posterior probability distribution for the parameters
  $m_l$ using two different data sets, CMB (solid) and CMB+BAO
  (dashed), normal and inverted ordering is designed by the
  blue and red curves respectively.}
\label{fig:1Dml}
\end{figure} 

\begin{figure}[ht!]
  \includegraphics[width=0.4\textwidth]{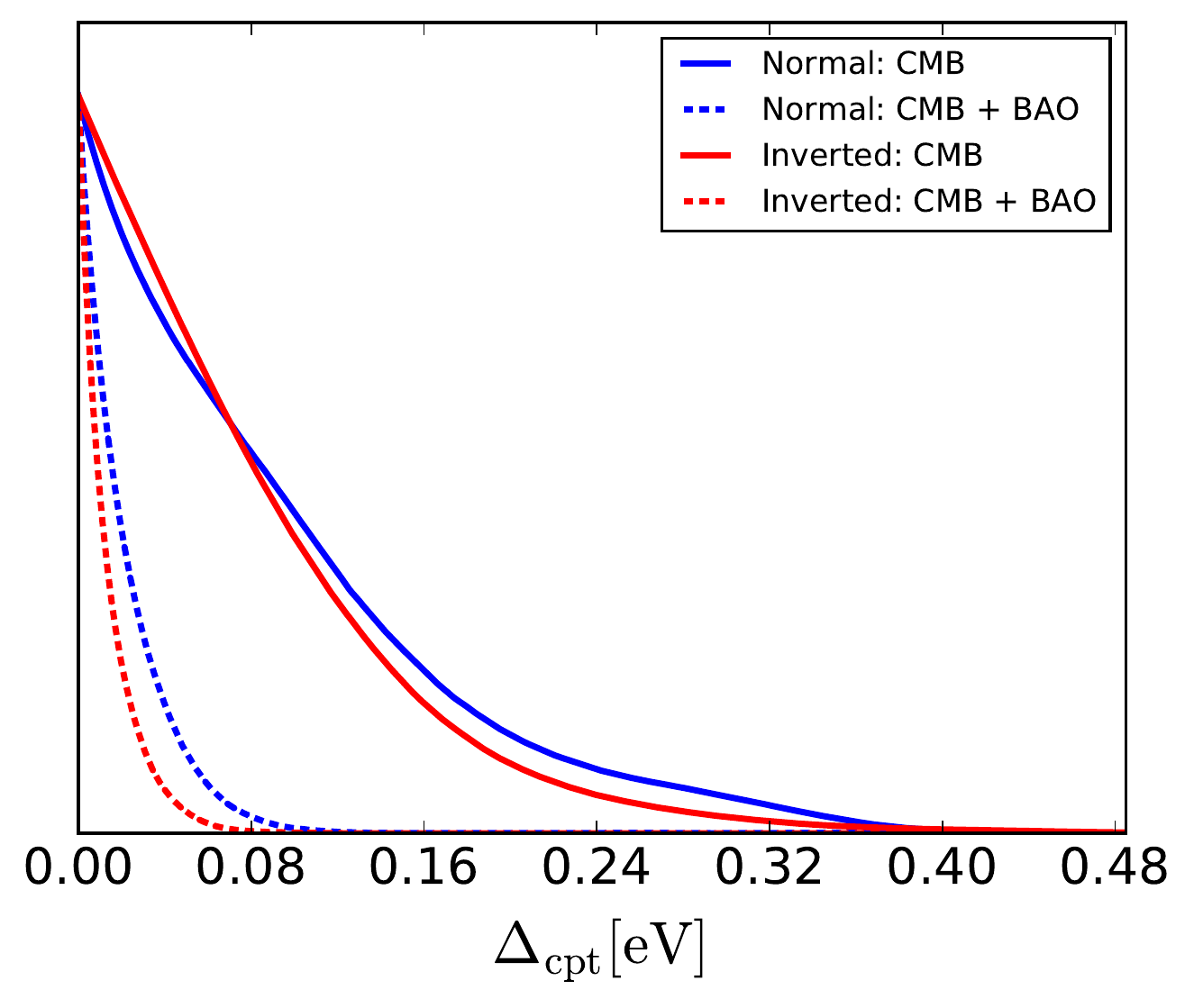}
\caption{1D posterior probability distribution for the parameters
  $\Delta_{CPT}$ using two different data sets, CMB (solid) and CMB+BAO
  (dashed), normal and inverted ordering is designed by the
  blue and red curves respectively.}
\label{fig:1Ddm}
\end{figure}

From the current cosmological data we use the combined (TTEEE,
low-l, lensing) data from Plank2015 \cite{Ade:2015xua} and the
measurement of the Baryonic Acoustic Oscillation (BAO) scale from
SDSS-DR10 SDSS-DR11 and 6dF\cite{Ross:2014qpa,Beutler:2011hx,Anderson:2013zyy}.
We do not include the less conservative local measurements of the local expansion rate
nor the full matter power spectrum since is more sensitive to the
treatment of the non-linear corrections and does not give a significant
improvement in the neutrino masses analysis\cite{Vagnozzi:2017ovm}.
In the following we designate by (CMB) the full set of Plank2015 data
and by (BAO) the combination of the Baryonic Acoustic Oscillation scale mentioned before.

The mean value and 95\% intervals for the two data sets CMB and
CMB+BAO and for the two cases, normal and inverted ordering are
summarized in tab.\ref{tab:results}.

\begin{table*}[t]
\def\arraystretch{1.2}
\begin{tabular}{|c|c|c|c|c|}
\hline
   \multirow{2}{2cm}{ Parameter} & \multicolumn{2}{c|}{Plank2015 (95\%)}  & \multicolumn{2}{c|}{Plank2015 + BAO (95\%) } \\
  \cline{2-5}
                               &\; Normal \; &\; Inverted\; & \;Normal\; & \;Inverted\;\\
  \hline
  \hline
  $10^{-2}\Omega_b h^2$ & $2.210^{+0.036}_{-0.034}$   &$2.206^{+0.032}_{-0.033}$ & $2.238^{+0.031}_{-0.034} $  &$2.240^{+0.028}_{-0.025}   $\\
  \hline
  $\Omega _{cdm} h^2$   & $0.1205^{+0.0031}_{-0.0030}$ &$0.1209^{+0.0030}_{-0.0027}$ &$0.1173^{+0.0022}_{-0.0025}$  &$0.1166^{+0.0021}_{-0.0021}$\\
  \hline
  $H0$                 & $63.7^{+2.6}_{-3.2}$       & $62.7^{+2.4}_{-3.0}$ &  $1.04203^{+0.00061}_{-0.00064}$  &  $65.97^{+0.99}_{-0.95}     $\\
  \hline
  $n_s$                & $0.9607^{+0.0095}_{-0.0094}$& $0.959^{+0.010}_{-0.010}$ & $0.9690^{+0.0092}_{-0.011} $  & $0.9716^{+0.0081}_{-0.0076}$ \\
  \hline
  $\log(10^{10}A_s)$    & $3.108^{+0.053}_{-0.053}$   & $3.117^{+0.055}_{-0.054}$ & $3.112^{+0.050}_{-0.052}   $  & $3.141^{+0.039}_{-0.038}   $ \\
  \hline
  $\tau_{reio }$        & $0.086^{+0.028}_{-0.029}   $ & $0.091^{+0.029}_{-0.029}$  &   $0.092^{+0.026}_{-0.027} $ & $0.107^{+0.021}_{-0.023}   $ \\
  \hline
  \multicolumn{5}{|c|}{Extended Parameters}\\
  \hline
  $m_l$ (eV)           & $< 0.133$                & $< 0.149 $ & $< 0.0491   $ &  $< 0.0423  $\\
  \hline
  $\Delta_{CPT}$ (eV)   & $< 0.255 $               & $< 0.215$ & $< 0.0588  $ &  $< 0.0428  $ \\
  \hline
\end{tabular}
\caption{Mean values and the 95\% regions for the
  parameters for normal and inverted ordering and for the different
  sets of cosmological data CMB and CMB+BAO.}
\label{tab:results}
\end{table*}

In fig.\ref{fig:1Dml} and fig.\ref{fig:1Ddm} we show the results of the posterior probability distribution for 
the new parameters $m_l$ and $\Delta_{CPT}$. For the sake of clarity and to make the comparison easier all the
one dimensional probability distributions are normalized such that they
get the same arbitrary value at the maximum.  

\begin{figure}[ht!]
    \includegraphics[width=0.4\textwidth]{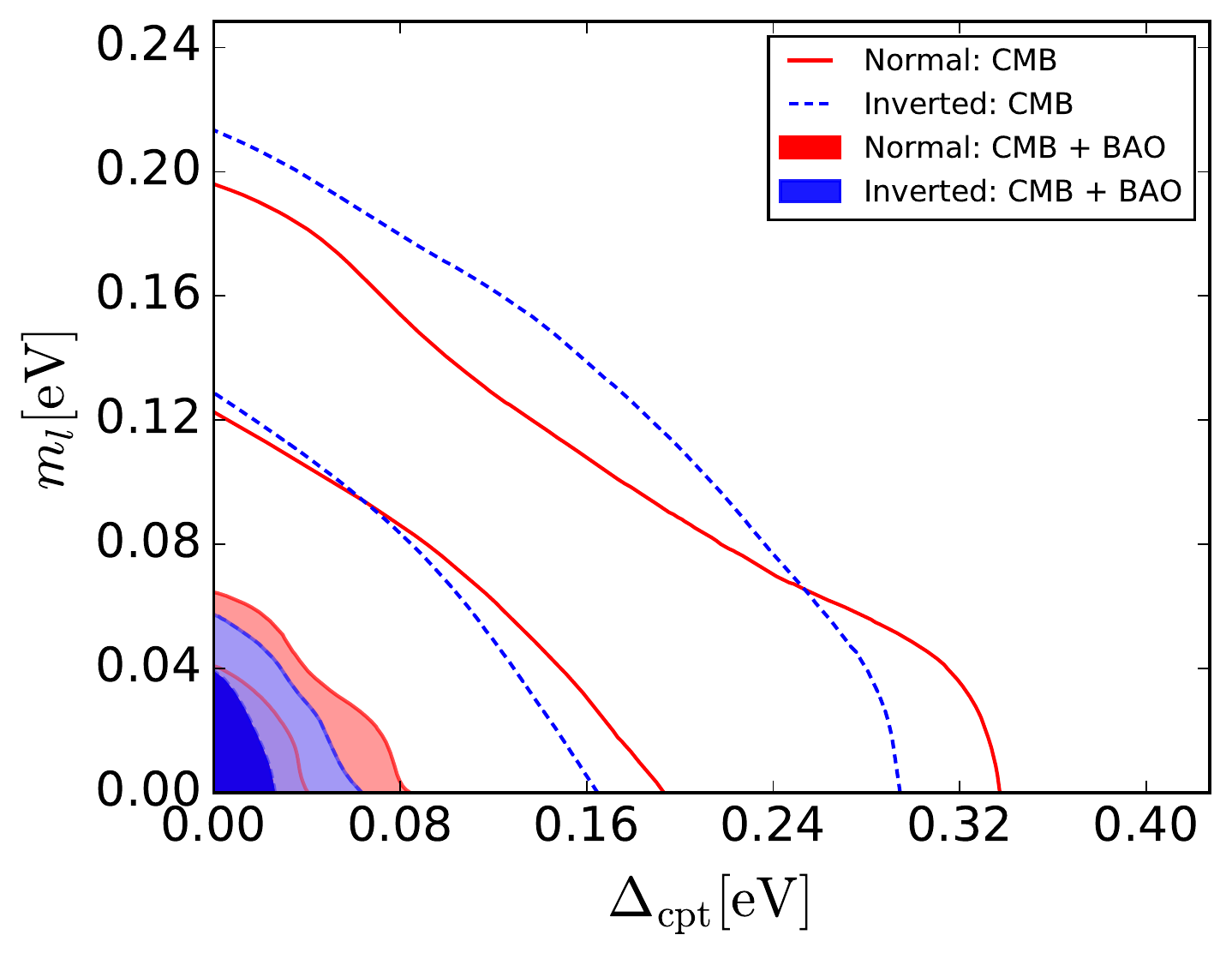}
\caption{68\% and 95\% probability contours for the light neutrino
  masses $m_l$ and $\Delta_{CPT}$, where (blue, red) and (solid, dashed) designate (normal, inverted) and (CMB, CMB+BAO) respectively.}
\label{fig:2D}
\end{figure}

In fig.\ref{fig:2D} we show the two dimensional 68\% and 95\%
probability contours for the following cases: normal ordering using CMB
only (red solid line), inverted ordering using CMB only (blue dashed line),
normal ordering with CMB+BAO (red filled contours) and inverted
ordering with CMB+BAO (blue filled regions). These constraints constitute the world best bound on CPT
violation in elementary particle masses so far.

In order to illustrate the potential of the near future data we perform an
analysis using a simulated power spectrum for Euclid, details can be
found in the Euclid Red Book\cite{Laureijs:2011gra}. 
For the forecast analysis we produce a simulated matter power spectrum data setting the
cosmological parameters to the $\Lambda CDM$ best fit and
$\Delta_{CPT}=m_l=0$. We do the forecast fit only for normal ordering.

\begin{figure}[ht!]
    \includegraphics[width=0.4\textwidth]{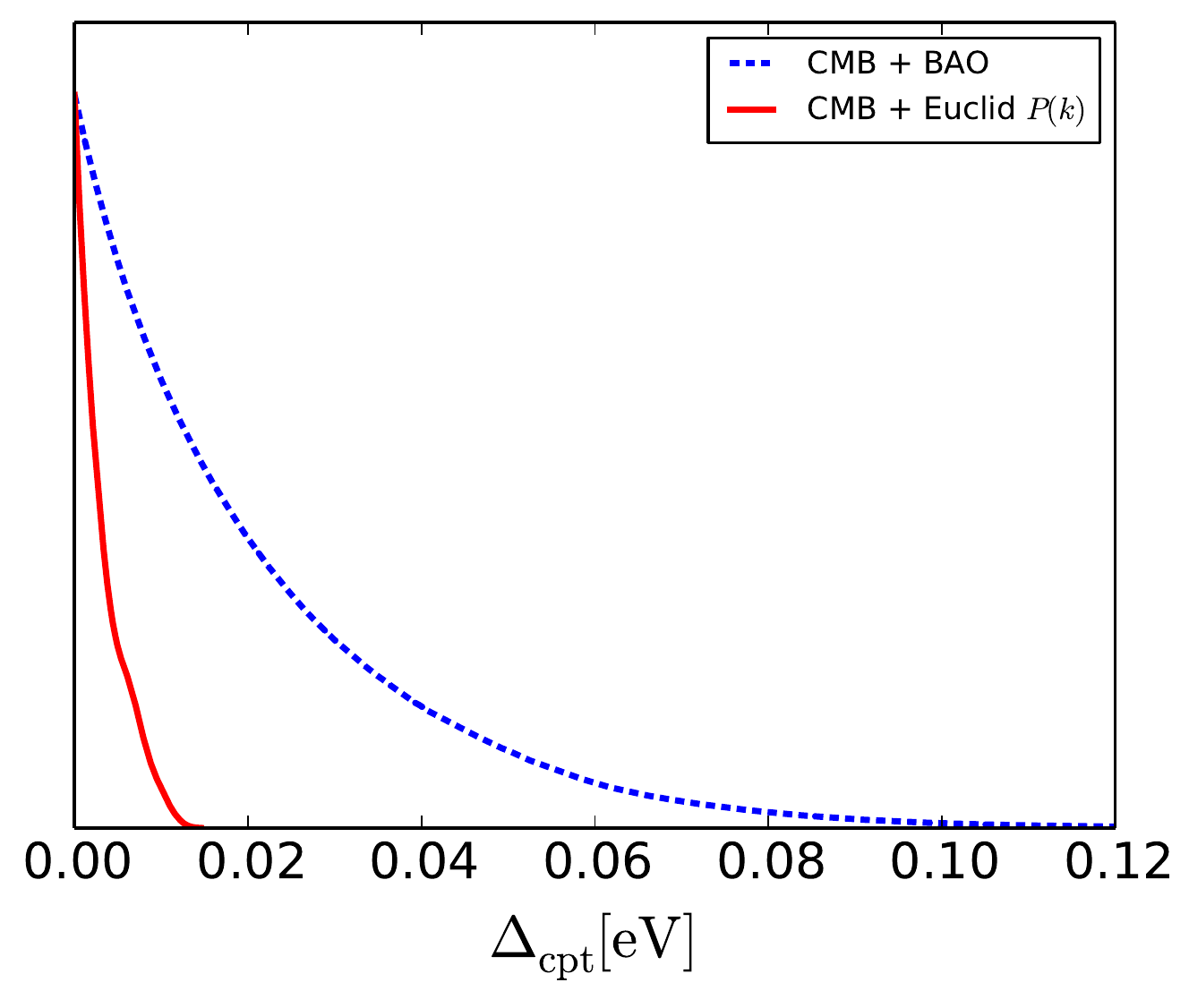} 
\caption{1D posterior probability distribution for the parameters
  $\Delta_{\rm CPT}$, the blue(dashed) is the most stringend bound
  with current data shown in fig.\ref{fig:1Ddm} and the red(solid) the
bound using generated Euclid power spectrum data with $\Delta_{CPT}=0$
and $m_l=0$.}
\label{fig:E1Ddm}
\end{figure}

The results of the forecast analysis compared with the most stringent
result using BAO measurements are shown in figure \ref{fig:E1Ddm}, for
the $\Delta_{CPT}$ parameter.
The results for the 95\% bound are $\Delta_{CPT} < 0.0088 {\rm eV}$ and $m_l < 0.02 {\rm eV}$. 

\section{Conclusions}
We give, for the first time, a bound on CPT violation in 
the absolute value of the neutrino-antineutrino mass splitting. Since the
kinematical laboratory experiments use only  antineutrinos they are not able
to give any bound on CPT, hence, for now, the only possibility to
bound $\Delta_{CPT}$ is to use cosmological data.

In order to do that we perform a full cosmological analysis using the
current CMB, and BAO data. Using only CMB the $95\%$ bounds are
$\Delta_{CPT}<0.26{\rm eV}$ and $\Delta_{CPT}<0.21{\rm eV}$  for
normal and inverted ordering respectively. Adding the BAO data we get
a more stringent bound, $\Delta_{CPT}<0.059{\rm eV}$ and
$\Delta_{CPT}<0.043{\rm eV}$ again for normal and inverted ordering respectively.

To illustrate the potential of the future data by Euclid satellite we
perform a forecast analysis where we generate a power spectrum for the
values $\Delta_{CPT}=0$ and $m_l=0$. Performing the fit together with
the Planck2015 data we get that the next generation of large scale
structure  experiments may give a 95\% bound to the CPT violation and
light neutrino masses of $\Delta_{CPT} < 0.0088 {\rm eV}$ and $m_l <
0.02 {\rm eV}$ respectively.

\section{Acknowledgements}
GB acknowledges support from the MEC and FEDER (EC) Grants
SEV-2014-0398, FIS2015-72245-EXP  and FPA2014-54459 and the
Generalitat Valenciana under grant PROMETEOII/2013/017. 
JS acknowledges support from FPA2014-57816-P and PROMETEOII/2014/050.
This project has received funding from the European Union Horizon 2020
research and innovation programme under the Marie Sklodowska-Curie grant
H2020-MSCA-ITN-2015//674896-ELUSIVES and and InvisiblesPlus
H2020-MSCA-RISE-2015, agreement No 690575.  
 

\end{document}